\documentclass{ws-ijmpd}
\usepackage[super,compress]{cite}
\usepackage{graphicx}
\usepackage{rotating}
\usepackage{booktabs}
\usepackage{longtable}
\usepackage{lscape}

\begin{document}

\markboth{X. Hernandez et al.}
{Wide binaries challenging a Newtonian prediction}

%
\catchline{}{}{}{}{}
%

\title{Challenging a Newtonian prediction through Gaia wide binaries}

\author{X. Hernandez}

\address{Instituto de Astronom\'{\i}a, Universidad Nacional Aut\'{o}noma de M\'{e}xico, \\
 Apartado Postal 70--264 C.P. 04510 \\ M\'exico D.F. M\'exico.\\
xavier@astro.unam.mx}

\author{R. A. M. Cort\'es}

\address{Instituto de Astronom\'{\i}a, Universidad Nacional Aut\'{o}noma de M\'{e}xico, \\
 Apartado Postal 70--264 C.P. 04510 \\ M\'exico D.F. M\'exico.\\
rcortes@astro.unam.mx}

\author{Christine Allen}

\address{Instituto de Astronom\'{\i}a, Universidad Nacional Aut\'{o}noma de M\'{e}xico, \\
 Apartado Postal 70--264 C.P. 04510 \\ M\'exico D.F. M\'exico.\\
chris@astro.unam.mx}

\author{R. Scarpa}

\address{Instituto de Astrof\'{\i}sica de Canarias, c/via Lactea s/n \\
San Cristobal de la Laguna 38205, Spain\\
Departamento de Astrof\'{\i}sica, Universidad de La Laguna (ULL)\\
38206 La Laguna, Tenerife, Spain.\\
riccardo.scarpa@gtc.iac.es}

\maketitle

\begin{history}
\received{Day Month Year}
\revised{Day Month Year}
\end{history}

\begin{abstract}
Under Newtonian dynamics, the relative motion of the components of a binary star should follow a Keplerian  
  scaling with separation. Once orientation effects and a
  distribution of ellipticities are accounted for, dynamical evolution can be modelled to include the effects
  of Galactic tides and stellar mass perturbers, over the lifetime of the solar neighbourhood. This
  furnishes a prediction for the relative velocity between the components of a binary and their
  projected separation. Taking a carefully selected small sample of 81 solar neighbourhood wide binaries from the 
  {\it Hipparcos} catalogue, we identify these same stars in the recent Gaia DR2, to test
  the prediction mentioned using the latest and most accurate astrometry available. The results are consistent
  with the Newtonian prediction for projected separations below 7000 AU, but inconsistent with it
  at larger separations, where accelerations are expected to be lower than the critical $a_{0}=1.2 \times
  10^{-10} $ { m s$^{-2}$} value of MONDian gravity. This result challenges Newtonian gravity at low
  accelerations and shows clearly the appearance of gravitational anomalies of the type usually attributed to dark
  matter at galactic scales, now at much smaller stellar scales.
\end{abstract}


\ccode{PACS numbers: 04.50.Kd, 95.35.+d, 97.10.Vm, 97.80.−d}


\section{Introduction}	

In galactic dynamics, the range of systems over which gravitational anomalies appear in the low acceleration regime extends across
vast astronomical scales. Ultra faint dwarf galaxies with scale radii of order a few tens of parsecs show stellar velocity dispersions
implying Newtonian mass to light ratios in the hundreds or even thousands (e.g. Ref.~\refcite{Ko15}). The classical dwarfs have sizes
of order of a kpc and mass to light ratios derived from observed stellar kinematics inconsistent
with those of naked stellar populations under standard gravity by well over an order of magnitude (e.g. Ref.~\refcite{Sa12}). This
reflects what is observed in spiral galaxies at tens of kpc, where rotation curves (e.g. Ref.~\refcite{Le17}) again yield
dynamics not corresponding to Newtonian dynamics given the empirically determined matter content. The trend has been recently extended to
include elliptical galaxies observed out to their external low acceleration regions by Ref.~\refcite{Du18}, and 
even for the case of Galactic globular clusters where velocity dispersion profiles suggest a change away from Newtonian dynamics
for low accelerations (e.g. Ref.~\refcite{Sc03}, Ref.~\refcite{Sc11}).

Empirically, the above gravitational anomalies can be described by MONDian dynamics (Ref.~\refcite{Mi84}, Ref.~\refcite{Sa02}),
where below an acceleration threshold of $a_{0}=1.2 \times 10^{-10}$ { m s$^{-2}$} kinematics stop falling along Newtonian expectations
of $v\propto R^{-1/2}$ to flatten out at the { Tully-Fisher} values of $V_{TF}=(G M a_{0})^{1/4}$ for centrifugal equilibrium velocities
or corresponding velocity dispersions for pressure supported systems, where $M$ is the total baryonic mass of the system in question.
The standard interpretation of {\bf these observations is the presence} of dominant halos of a yet undetected hypothetical dark matter
component surrounding the astrophysical systems being observed.

Wide binary pairs in the solar neighbourhood offer an opportunity to probe dynamics at low accelerations on the smallest
astrophysical scales. In principle these can yield crucial restrictions on the structure of gravity at low accelerations and lengths
where the presence of dark matter is not expected. For a solar mass binary, at separations of above $3.4\times 10^{-2}$pc,
$7000$ AU, accelerations will fall below $a_{0}$ under Newtonian expectations. A first attempt in this direction was made by
two of us in Ref.~\refcite{He12} where we used the Ref.~\refcite{Sh11} -henceforth SO11- carefully selected sample of
{\it Hipparcos} wide binaries. This catalogue includes a full Bayesian model and use of local 5-dimensional phase space
density to identify wide binary candidates and rigorously assign a probability that each candidate forms a physical system,
rather than being the result of chance associations.

Retaining a sample of binaries from SO11 where contamination was limited to less than $10\%$,
in  Ref.~\refcite{He12} results indicated relative velocities for the binary pairs studied above the Newtonian
expectations for accelerations below $a_{0}$. This remained true even after accounting for projection effects, ellipticity distributions
and disruption and evolution of ionised binaries due to the Galactic tidal field and encounters with field stars and stellar remnants,
as modelled by  Ref.~\refcite{Ji10}, albeit the relatively large errors in proper motions present in the {\it Hipparcos} catalogue.

One of us, in  Ref.~\refcite{Sc17}, explored the problem using a small sample of 60 candidate wide binaries with projected
separations between 0.004 and 1.0 pc. That study found that a number of wide binaries are capable of surviving the galactic tides and
stellar encounters of the solar neighbourhood dynamical environment, with a small sub-sample of the widest pairs showing kinematics
more consistent with MONDian dynamics than Newtonian ones. More recently, theoretical studies by Ref.~\refcite{Pi18}
and Ref.~\refcite{Ban18} confirmed that Gaia data, in terms of expected number of detected wide binaries and confidence
intervals for the relevant proper motions and parallaxes, are sufficient to detect MONDian deviations from Newtonian
dynamics in the low acceleration regime probed by these systems, should they be present.

The obvious next step is to reproduce the careful and detailed procedure presented in SO11, but using
this time the Gaia DR2 catalogue. This painstaking approach will ultimately furnish a definitive answer regarding the
presence of gravitational anomalies at stellar scales in the low acceleration regime, but is currently
hampered by our incomplete understanding of the problems still
present in the data of the very novel Gaia DR2. For example, only about two
thirds of the Hipparcos2 sources have Gaia DR2 counterparts (Ref.~\refcite{Ma18}). Also, Gaia treats all binaries closer
than about 1" (depending on the magnitude difference) as single sources, which may give anomalous parallaxes, and the
parallax solutions may be more sensitive to duplicity in certain areas, etc. (see
Ref.~\refcite{Ga18} A1 for some of the known issues.)

A first order sampling of the issue can be more directly probed by taking advantage of the correspondence between the
Gaia and the {\it Hipparcos} catalogues. One can take the sample selection from the accurate Bayesian analysis of SO11,
and the actual astrometric data from the Gaia DR2. Here we present results of such an approach, yielding
a small sample of 81 wide binaries from the original SO11 catalogue having a $<10\%$ probability of being chance
associations, Gaia DR2 data consistent with the original {\it Hipparcos} reported quantities, and consistent parallaxes for both
components in the Gaia DR2. Although it is only a reduced sample,
the superior quality of the Gaia satellite allows us to infer relative velocities for the binaries in question
to a much higher degree of accuracy than what was available to Ref.~\refcite{He12}. Interestingly, our results
show a departure from Newtonian predictions as projected separations grow beyond the critical
0.034 pc. Indeed, our new results are consistent with what was reported in Ref.~\refcite{He12}: the mean
values of the inferred relative velocities are essentially unchanged, with the error bars showing a dramatic reduction.
This effectively rules out the Newtonian prediction of Ref.~\refcite{Ji10} and provides solid evidence for a gravitational
anomaly in the low acceleration regime, this time at stellar scales.

\section{Sample selection}

As outlined in the introduction, the sample selection is based on the wide binary catalogue of SO11,
which was constructed using a very detailed Bayesian procedure. This identifies and quantifies the probability of each binary pair
being an actual physical system, rather than the result of projection effects or chance associations, including also the
Tycho-2 and the {\it Tycho} double star catalogues (Ref.~\refcite{Ho00} and Ref.~\refcite{Fa02}). To that end a 5-dimensional
space of spatial positions and proper motions was cross-correlated with a galactic phase-space density library, explicitly
excluding the largest known local star clusters, to identify binary candidates as significant local over-densities in phase
space. Corrections due to spherical projections effects were also considered, to yield a catalogue of 840 wide binaries with projected
separations of between 0.003 and 10 pc and, crucially, a well determined probability of chance association, $P_{ch}$. Taking only those
pairs where this probability satisfies $P_{ch}<10 \%$, reduces the original SO11 sample to 359 wide binaries. This catalogue is also
narrowly restricted in spectral type for both primaries and companions of each binary, yielding stars in a narrow range of masses centred
on $0.5 M_{\odot}$. This last is important as it allows a clean comparison to the fixed mass binary simulations of Ref.~\refcite{Ji10},
see below. Following the original SO11 terminology, the brightest star in each binary system is termed the primary.

The SO11 search criteria ensure the absence of near neighbours, and result in binary candidates with separations
which are always many times smaller than the typical interstellar separations at the location of the binaries in question. 
Extensive testing with synthetic samples in SO11 guarantees the catalogue includes very few multiple systems with
undetected extra companions and is highly complete in the 6 to 100 pc distance range from the Sun.

We now take advantage of the {\it Hipparcos} to Gaia DR2 correspondence availability to search for the updated astrometry of
the 359 $P_{ch}<10\%$ wide binary SO11 sample in the Gaia DR2. The search returns only 151 pairs where each component
of the SO11 binaries appears in the Gaia DR2 with two proper motion parameters and measured parallax.
This is not surprising, since only about two thirds of the Hipparcos2 sources have
Gaia DR2 counterparts (Ref.~\refcite{Ma18}). It is not yet clear why there are so many
sources missing. According to the above authors a combination of effects may be
present. As each binary is excluded if either component is absent from the DR2, the fraction we obtain is typical.
Next, the SO11 catalogue returns a few systems where more than one secondary is identified as the companion to a given
primary, and also cases where a single secondary is identified as the companion to more than one primary. We remove all these
cases of multiple identifications, bringing the sample down to 131 binaries. 

\begin{figure}[pb]
\centerline{\psfig{file=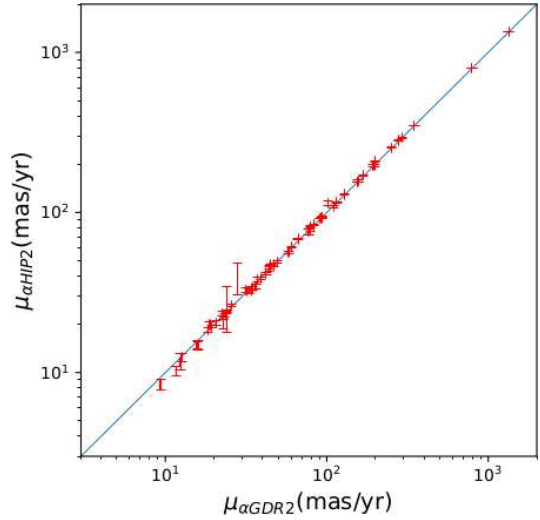,width=7cm}}
\vspace*{8pt}
\caption{Comparison of SO11 {\it Hipparcos} and Gaia DR2 proper motion data in right ascension for each of the
  two components of the binaries studied. The agreement with the identity line in the majority of cases shows the stars in question
  have been successfully identified from the first catalogue to the Gaia DR2 sample.}
\end{figure}

\begin{figure}[pb]
\centerline{\psfig{file=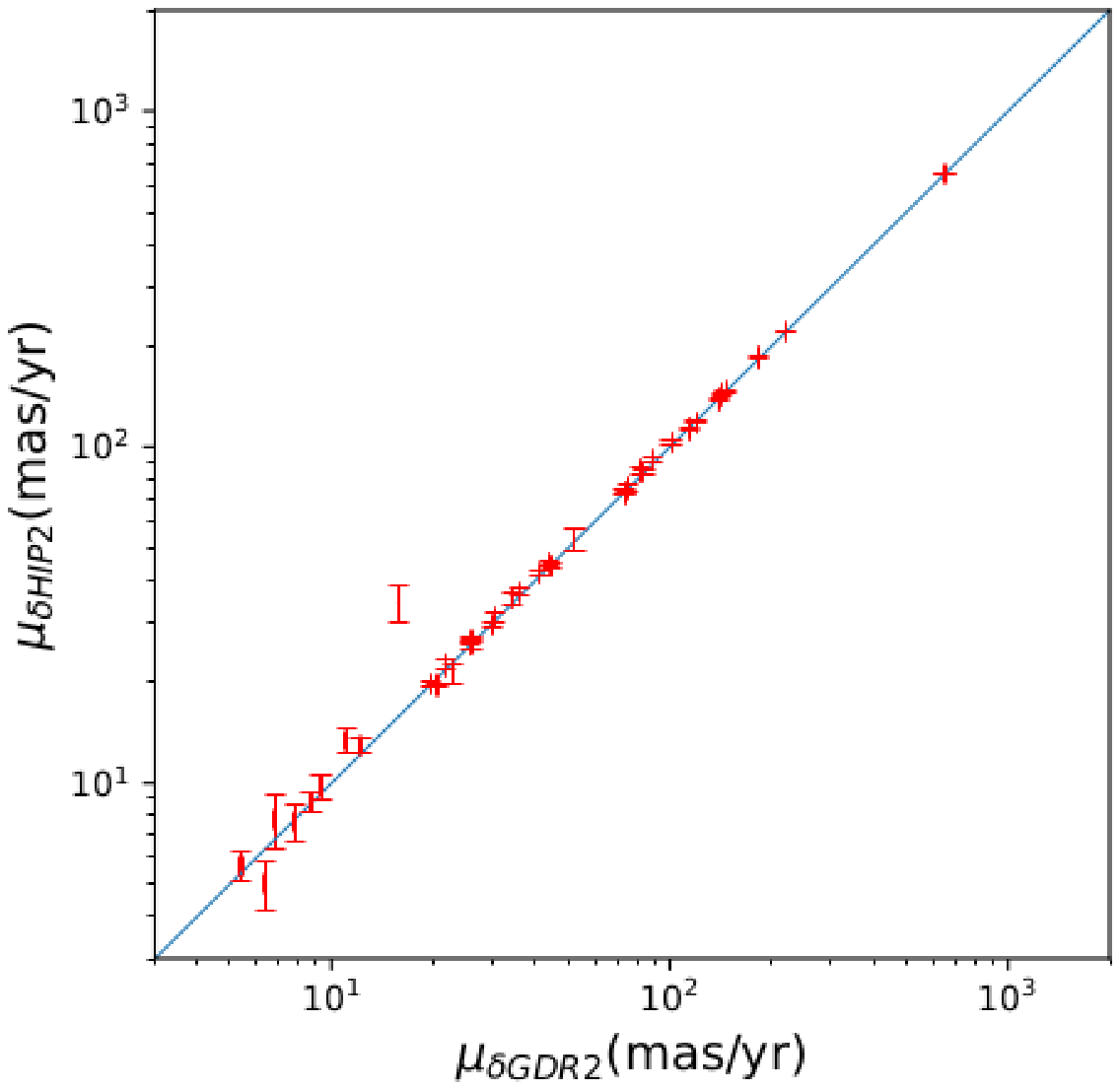,width=7cm}}
\vspace*{8pt}
\caption{Comparison of SO11 {\it Hipparcos} and Gaia DR2 proper motion data in declination for each of the
  two components of the binaries studied. The agreement with the identity line in the majority of cases shows the stars in question
  have been successfully identified from the first catalogue to the Gaia DR2 sample.}
\end{figure}

\begin{figure}[pb]
\centerline{\psfig{file=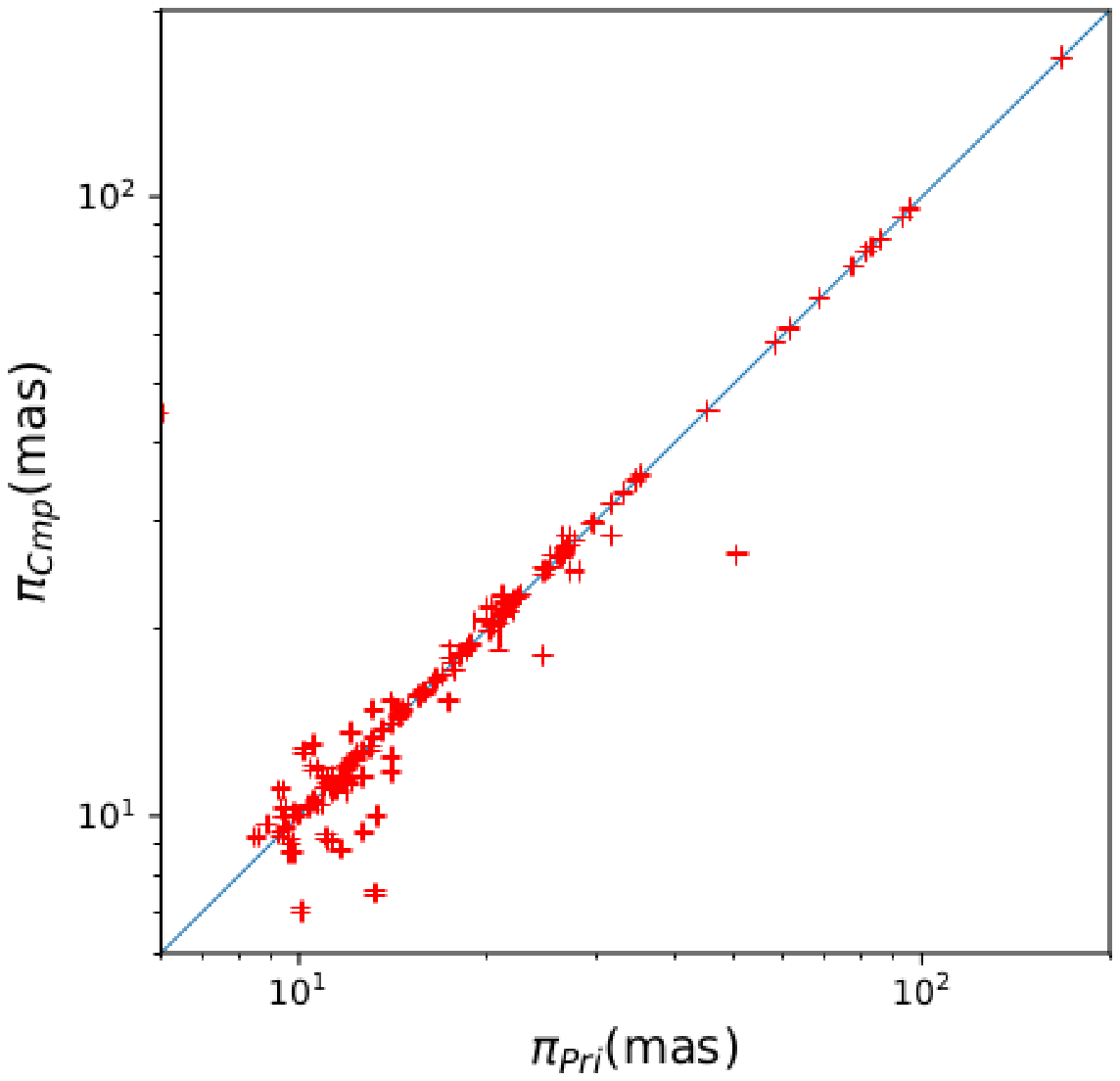,width=7cm}}
\vspace*{8pt}
\caption{Comparison of the Gaia DR2 reported parallax for the two distinct components of each of the binaries studied. The
  disagreement of only a handful of cases with the identity line allows to exclude such discordant pairs as part of the
  expected $10\%$ misidentified binaries in the original SO11 catalogue, or as misidentified stars in
  going from the {\it Hipparcos} catalogue to the Gaia DR2.}
\end{figure}

A first test of the reliability with which the {\it Hipparcos} binaries have been identified in the Gaia DR2 comes from
comparing the proper motion measurements reported in SO11 with the corresponding measurements
reported in the Gaia DR2. This is shown in Figures 1 and 2, where the ranges shown in the axes were chosen so as to display
clearly most of our sample; a handful of very discordant systems do not appear in the plots, as they are very far from
the identity line shown. It is clear that in most cases, the proper motion values reported by both satellites are in agreement
with each other, to within their respective confidence intervals, the {\it Hipparcos} error ranges being much larger than the
Gaia DR2 ones. Still, we introduce a cut to remove from consideration any binary candidate where for any component the
{\it Hipparcos} and Gaia DR2 {\bf proper motions} are more than $3\sigma$ from each other. Our final results are not sensitive
to this threshold, provided the few discordant misidentified binaries are removed. This cut leaves us with only 117 stellar pairs.
Next, as shown in Figure 3, we check that the Gaia DR2 reported parallax measurements for each of the primaries and companions are
not discordant, and remove from the sample any binary where the distances to each component are not within 13\% of each other.
This exclusion criterion identifies 17 candidates, thus reducing the sample to 100 pairs. Notice that the
test shown in Figures 1 and 2 effectively gives us a 10 year baseline which, on top of the robustness in the SO11 catalogue
towards multiple systems, where no radial velocities are involved, removes any remaining binary where a third component might
be altering proper motions with timescales shorter than 10 years (see Ref.~\refcite{Ban18}).

Finally, we take advantage of the Gaia radial velocity measurements (when available) and remove any binaries where the radial
velocity difference, $\Delta V_{r}$, between both components is larger than $4$ km s$^{-1}$. The resulting cut is not very
sensitive to this velocity threshold, as the removed binary pairs typically have much larger and discordant $\Delta V_{r}$ values,
with an average value for those removed of $\Delta V_{r}=28$ km s$^{-1}$. Our final sample comprises 81 binary pairs.

Thus, we have prefered very strict cuts to our final sample which leave us with modest numbers, but guarantee the exclusion
of misidentified stars in going from {\it Hipparcos} data to Gaia DR2 and chance alignment contamination in the original
SO11 catalogue, all of which become conspicuous in the comparisons presented in this section. Table 1 summarises the Gaia DR2
properties used and catalogue numbers from both {\it Hipparcos} and Gaia for the primaries and companions of all the binaries used,
together with the results of the exclusion criteria described.

\section{Gaia wide binary projected kinematic results}

\begin{figure}[pb]
\centerline{\psfig{file=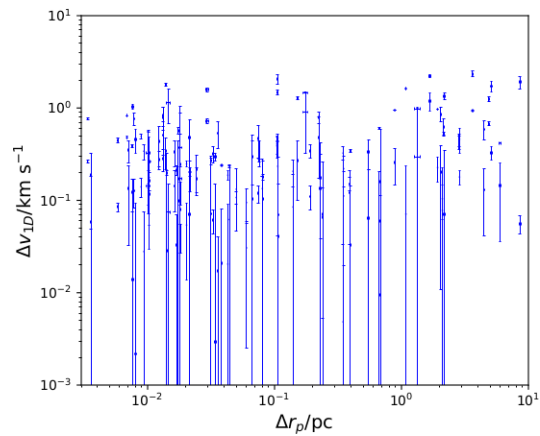,width=7cm}}
\vspace*{8pt}
\caption{One dimensional relative velocities for the final binary sample, showing both results from using only right
  ascension data, and from using only declination values, with corresponding error bars.}
\end{figure}

\begin{figure}[pb]
\centerline{\psfig{file=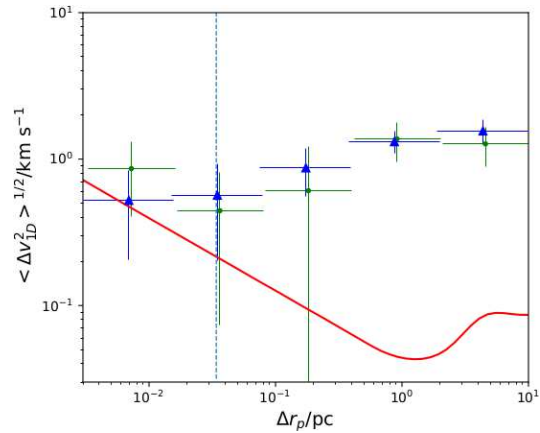,width=7cm}}
\vspace*{8pt}
\caption{The solid curve shows the rms. value for the one dimensional relative velocity between the components of a present
  day Solar Neighbourhood binary as a function of projected separation, for the Newtonian prediction of Ref. \citen{Ji10}.
  The same quantity for the SO11 wide binary catalogue stars, this time using Gaia DR2 data, is shown by the
  points with error bars, circles for right ascension data and triangles for declination. A small artificial horizontal offset
  was introduced on both data sets to avoid having overlapping error bars. The inconsistency of the observed
  data with the Newtonian prediction is obvious, for separations greater than the 7000 AU at which accelerations are expected to
  drop below {$a_{0}$}. This threshold is indicated by the dashed vertical line.}
\end{figure}

In Ref.~\refcite{He12} we calculated the projected separation in the plane of the sky using only the parallax to the
primary of each binary, but given the higher quality Gaia DR2 data, we now compute the projected separation between the
components of each binary using explicitly the observed Gaia positions and parallaxes to each component of the binary.
The average parallax to both components is used to gauge the distance to each pair. Using reported Gaia proper motions,
the relative velocity difference in one dimension is calculated for each binary twice, once considering only right ascension
proper motions, and once considering only declination proper motions. In all cases the relative physical motion is calculated
including full spherical geometric effects (e.g. Ref.~\refcite{Sm68}), and not under the more standard small angle approximation.
This requires the radial velocity of at least one component, which we have for 71 of our 81 final pairs. For the remaining
10, a $V_{r}=0$ was assumed for the effects of the above correction (e.g. SO11). The effects of the above refinement are only
relevant for the nearest and widest of pairs, in our sample only a few individual systems, as can be {\bf readily checked}
from the data used and shown in Table 1.

Figure 4 gives the two $\Delta v_{1D}$ measurements for each binary pair in the final sample, with corresponding $1\sigma$
error bars. A clear flat upper envelope is evident. In Figure 5 we show the rms. value for the one dimensional velocity
differences described above, plotted against projected separations in a binned logarithmic scale, circles and triangles for
right ascension and declination data, respectively. The horizontal error bars give the bin sizes, while the vertical ones
show the contribution of Gaia reported errors and error propagation, to which a Poisson contribution has been added, and
which given the small numbers of binaries in each bin (21, 23, 17, 8 and 12, from left to right), actually dominates the
error budget in most cases. The dashed vertical line appears at 7000 AU, the approximate scale where acceleration is expected
to drop below $a_{0}$.

Also shown in Figure 5 are the Newtonian predictions for this same quantity from Ref.~\refcite{Ji10}, where large
collections of 50,000 simulated binaries are modelled for a range of plausible distributions of ellipticities, and
followed dynamically under Newtonian expectations within the local Galactic tidal field. These are also subject to the effects of
field star and field stellar remnant bombardment for a 10 Gyr period. Finally, the resulting bound and un-bound stellar
pairs are projected along a fiducial line of sight to yield a robust prediction for the expected $<\Delta v^{2}_{1D}>^{1/2}$
as a function of $\Delta r$, solid curve. This results are easy to understand; a $\Delta v \propto \Delta r^{-1/2}$ trend
is apparent, down to the tidal radius of the problem which appears at 1.7 pc. Beyond this point, ionised binaries
continue to move along practically common Galactic orbits, with relative velocities which show a mild enhancement which
then levels off at close to $0.1$ km s$^{-1}$.

It is clear that to the left of the $a_{0}$ dashed line, our results are consistent within confidence intervals with the
Newtonian expectations. However, going to separations larger than 7000AU, the observed points stop following the expected
trend and actually level off at a $\Delta v$ amplitude close to the values seen at the $a_{0}$ point, reproducing qualitatively
the phenomenology seen in galactic rotation curves.

This result for the binary sample presented is extremely challenging to a Newtonian point of view, where the relative
velocities are expected to be much lower than observed. Given the construction of the SO11 sample, binaries with small
velocity differences would appear as stronger local over-densities in phase space, and hence, selection criteria, if anything,
are biased against binaries with large velocity differences, not small ones. Thus, from a Newtonian point of view, bound binaries
with separations smaller that the tidal radius of 1.7 pc and larger than 7000 AU are unexpectedly missing. Also, as already
mentioned in Ref.~\refcite{Oh17}, a population
of non-chance associated binaries appears at scales above 7000AU having relative velocities over an order of magnitude above
bound expectations, { with relative velocities of $\sim$ 1 km s$^{-1}$ and separations of a few pc; the dynamical ages of
un-bound systems are of only a few $10^6$ years.} What sustains and replenishes these populations under a Newtonian framework?
At separations below the 1.7 pc tidal radius of the problem, bound binaries should appear, under a Newtonian framework.

Furthermore, the results shown in figure 5 confirm what was presented in  Ref.~\refcite{He12}, though
the much coarser {\it Hipparcos} data of that first study yielded significantly larger error bars. That those first
results might have been the result of missed biases or simply the error structure of the {\it Hipparcos} data now appears
very unlikely, as we see two consistent results coming from data obtained by two completely independent satellites. Indeed,
the error bars have significantly shrunk, with central results changing little. Note also that the two $\Delta v_{1D}$
estimates we obtain, using only right ascension or only declination data, are consistent with each other.

A potential caveat on the interpretation presented comes from the possible  presence of a population of misidentified
binaries being actually parts of loose, dissolving moving groups.  Ref.~\refcite{Oh17} recently showed that although isolated
wide binaries dominate, a search algorithm of the type used in SO11 could also pick up a fraction of misidentified
binaries being part of larger moving groups, many of the smallest of which probably remain undiscovered. A full answer
and validation or otherwise of our results, necessarily awaits a much more extensive study through the full Gaia catalogue,
and a more complete exploration and understanding of the phase-space structure and over-densities of the Solar Neighbourhood and
local Milky Way disk environment.

Although the gravitational anomaly detected appears on crossing the $a_{0}$ threshold of MOND, in MOND as such, the results
are equally unexpected as the external field effect of MOND (or AQUAL e.g.  Ref.~\refcite{Sa02} should dominate. Given
that the
orbital acceleration of the solar neighbourhood is higher than the internal acceleration of the binaries in question, in
MOND as such, only a very modest enhancement of the effective value of G would be expected (e.g.  Ref.~\refcite{Pi18}).
Thus, within a MONDian frame our results imply the validity of not the most well studied version, but rather of a variant where the external
field effect does not appear, or is much suppressed e.g. as in  Ref.~\refcite{Mi11}. In terms of covariant extensions to GR having a
MONDian low velocity limit, it is hard to know to what extent an external field effect might be present in many of the recently
developed options (e.g. the f(R) variants reviewed in Ref.~\refcite{Ca11}, the emergent gravity of Ref.~\refcite{Ve17}
or the F(R,L) proposal of Ref.~\refcite{Bar18}. Our results then serve as constraints in terms of requiring a
minimal external field effect, at least for the sub-parsec scales in the solar neighbourhood explored.

\section{Final remarks}

We have presented a sample of 81 wide binaries which were very carefully selected against chance associations or projection
effects through the cross correlation of the {\it Hipparcos}, Tycho-2 and the {\it Tycho} double star catalogues, amongst others,
with the detailed 5-dimensional phase space structure of the solar neighbourhood by SO11. By taking advantage
of the cross-identification of the {\it Hipparcos} catalogue and the Gaia DR2 data, we updated the parallax and proper motion
observations of SO11 to use exclusively Gaia DR2 astrometry.

These binaries were then compared to Newtonian predictions for the expected one dimensional rms. relative velocity between the
components of each binary and their projected separations, including modeling orientation effects, a number of plausible
distributions of ellipticities and, crucially, the effects of Galactic tides and stellar and stellar remnant perturbers over a
10 Gyr period, by Jiang \& Tremaine (2010).

For separations below 7000 AU, where accelerations are expected to be above the $a_{0}=1.2\times 10^{-10}$ { m s$^{-2}$} of MOND,
we find the data to be consistent with the Newtonian predictions. For projected separations above 7000 AU, however, the
data are inconsistent with Newtonian predictions. This challenges the validity of Newtonian dynamics at
the low acceleration regime, and shows the existence of gravitational anomalies of the type generally attributed to the
presence of a hypothetical and dominant dark matter component, this time down to the relatively tiny sub-parsec stellar scales.

\tiny
\begin{landscape}
\begin{longtable}{rrrrrrrrrrl}
  \toprule
      Index &    HIP2 & \multicolumn{1}{c}{GDR2} & \multicolumn{1}{c}{$\alpha$} & \multicolumn{1}{c}{$\delta$} & \multicolumn{1}{c}{$\mu_\alpha$} & \multicolumn{1}{c}{$\mu_\delta$} & \multicolumn{1}{c}{$v_r$} & \multicolumn{1}{c}{$d$} & \multicolumn{1}{c}{$\log_{10}(\Delta R)$} & Exclusion\\
 \cmidrule(lr){6-7} \cmidrule(lr){9-10}
\multicolumn{1}{c}{SO11}  &  & & \multicolumn{2}{c}{(deg)} & \multicolumn{2}{c}{(mas/yr)} & \multicolumn{1}{c}{(km/s)} & \multicolumn{2}{c}{(pc)}& \multicolumn{1}{c}{Test} \\
\midrule
\endhead
\midrule
\multicolumn{11}{r}{{Continued on next page}} \\
\midrule
\endfoot

\bottomrule
\\
\caption{ \normalsize {\it Hipparcos} and Gaia identification numbers for the 133 binary stars used, together with Gaia DR2 data and
  results of the
      exclusion tests performed: (a) shows an individual star excluded (which results in the exclusion of the respective binary) for
      having an inconsistent proper motion in right ascension between {\it Hipparcos} and the DR2, (b) the corresponding result for
      declination, (c) shows a binary excluded for having discordant parallaxes to both components (distance differences larger than
      13 \%) in the DR2 data, and (d) gives binaries removed for having DR2 radial velocity differences $>4$ km s$^{-1}$, with the
      81 retained binaries appearing blank in this column. Upper rows give data of primaries and lower ones of secondaries.}

\endlastfoot 
  \hline 16 &   15371 &      4722135642226356736 &                    49.565826 &                   -62.503574 &               1331.151$\pm$0.355 &                648.523$\pm$0.431 &          12.01$\pm$  0.32 &        12.046$\pm$0.027 &                                           &                                                    \\
            &   15330 &      4722111590409480064 &                    49.454851 &                   -62.572523 &               1337.591$\pm$0.142 &                649.930$\pm$0.154 &          12.21$\pm$  0.17 &        12.039$\pm$0.011 &                          -1.744$\pm$0.002 &                                                    \\
  \hline 17 &   17414 &        43335880716390784 &                    55.969727 &                    16.670667 &                157.945$\pm$0.086 &               -316.326$\pm$0.052 &          34.18$\pm$  0.15 &        17.209$\pm$0.012 &                                           &                                                    \\
            &   17405 &        43335537119008896 &                    55.939232 &                    16.665933 &                156.215$\pm$0.089 &               -310.291$\pm$0.064 &          34.22$\pm$  1.08 &        17.239$\pm$0.014 &                          -2.051$\pm$0.001 &                                                    \\
  \hline 21 &   19859 &      3285218186904332288 &                    63.869527 &                     6.186399 &               -109.700$\pm$0.095 &               -107.368$\pm$0.079 &          -7.21$\pm$  0.15 &        22.087$\pm$0.026 &                                           &                                                    \\
\end{longtable}
\end{landscape}

\normalsize

\section*{Acknowledgments}

The authors thank an anonymous referee for a careful reading of the first version of this paper and constructive
criticism leading to a more complete and clearer final version.
XH and RAMC acknowledge the support of DGAPA-UNAM PAPIIT IN-104517 and CONACyT.



\end{document}